\documentclass[11pt]{article}
\usepackage{amsmath,amssymb,amscd,amsthm}
\usepackage{epsfig}

\newtheorem{theorem}{Theorem}
\newtheorem{lemma}{Lemma}
\newtheorem{proposition}{Proposition}
\newtheorem{remark}{Remark}

\newtheorem{definition}{Definition}
\newtheorem{corollary}{Corollary}

\newtheorem{claim}{Claim}

\newcommand{\f}[2]{\frac{#1}{#2}}

\newcommand{\De}{\Delta}
\newcommand{\ve}{\varepsilon}

\newcommand{\la}{\lambda}

\newcommand{\si}{\sigma}

\newcommand{\om}{\omega}

\newcommand{\cf}{\mathcal F}
\newcommand{\cb}{\mathcal B}

\newcommand{\suml}{\sum\limits}

\newcommand{\beq}{\begin{equation}}
\newcommand{\eeq}{\end{equation}}
\newcommand{\beqna}{\begin{eqnarray*}}
\newcommand{\eeqna}{\end{eqnarray*}}
\newcommand{\beqn}{\begin{equation*}}
\newcommand{\eeqn}{\end{equation*}}
\newcommand{\bp}{\begin{proof}}
\newcommand{\ep}{\end{proof}}
\newcommand{\bprop}{\begin{proposition}}
\newcommand{\eprop}{\end{proposition}}
\newcommand{\bt}{\begin{theorem}}
\newcommand{\et}{\end{theorem}}
\newcommand{\bex}{\begin{Example}}
\newcommand{\eex}{\end{Example}}
\newcommand{\bc}{\begin{corollary}}
\newcommand{\ec}{\end{corollary}}
\newcommand{\bcl}{\begin{claim}}
\newcommand{\ecl}{\end{claim}}
\newcommand{\bl}{\begin{lemma}}
\newcommand{\el}{\end{lemma}}

{\begin{trivlist} \item[]{\em Proof }}%
{\hspace*{\fill}$\rule{.3\baselineskip}{.35\baselineskip}$\end{trivlist}}

\begin{document}

\title{\bf On the spectral theory and dispersive estimates \\
for a discrete Schr\"{o}dinger equation in one dimension}
\author{D.E. Pelinovsky$^1$ and A. Stefanov$^2$ \\
{\small $^{1}$ Department of Mathematics, McMaster University,
Hamilton, Ontario, Canada, L8S 4K1} \\
{\small $^{2}$ Department of Mathematics, University of Kansas,
1460 Jayhawk Blvd, Lawrence, KS 66045--7523} }

\maketitle

\begin{abstract}
Based on the recent work \cite{KKK} for compact potentials, we
develop the spectral theory for the one-dimensional discrete
Schr\"odinger operator
$$
H \phi = (-\De + V)\phi=-(\phi_{n+1} + \phi_{n-1} - 2 \phi_n) + V_n \phi_n.
$$
We show that under appropriate decay conditions on the general potential
(and a non-resonance condition at the spectral edges), the spectrum of $H$ consists of finitely
many eigenvalues of finite multiplicities and the essential (absolutely continuous)
spectrum, while the resolvent satisfies the limiting absorption principle and the
Puiseux expansions near the edges. These properties imply the dispersive estimates
$$
\|e^{i t H} P_{\rm a.c.}(H)\|_{l^2_{\sigma} \to l^2_{-\sigma}} \lesssim t^{-3/2}
$$
for any fixed $\sigma > \frac{5}{2}$ and any $t > 0$, where $P_{\rm a.c.}(H)$
denotes the spectral projection to the absolutely continuous spectrum of $H$.
In addition, based on the scattering theory for the discrete Jost solutions
and the previous results in \cite{SK}, we find new dispersive estimates
$$
\|e^{i t H} P_{\rm a.c.}(H) \|_{l^1\to l^\infty}\lesssim t^{-1/3}.
$$
These estimates are sharp for the discrete Schr\"{o}dinger operators
even for $V = 0$.
\end{abstract}

\section{Introduction}

We consider a stationary one-dimensional discrete Schr\"{o}dinger equation in the
form
\begin{equation}
\label{statdNLS} H \phi:=(-\Delta + V) \phi = \lambda \phi \quad
\Leftrightarrow \quad -(\phi_{n+1} + \phi_{n-1} - 2 \phi_n) + V_n
\phi_n = \lambda \phi_n,
\end{equation}
where $V$ is a real-valued potential on $\mathbb{Z}$, $\lambda \in
\mathbb{C}$ is a spectral parameter, and $\phi$ is an
eigenfunction in an appropriate space. We will use plain letters
$V$ and $\phi$ to denote sequences $\{ V_n \}_{n \in \mathbb{Z}}$
and $\{ \phi_n \}_{n \in \mathbb{Z}}$. We will also use standard weighted
spaces $l^2_{\si}$ and $l^2_{\si}$ on $\mathbb{Z}$ equipped with the norms
\begin{equation}
\| u \|_{l^2_\si} = \left( \suml_{n\in \mathbb{Z}} (1+n^2)^{\si}
|u_n|^2 \right)^{1/2}, \qquad \| u \|_{l^1_\si} = \suml_{n\in \mathbb{Z}} (1+n^2)^{\si/2}
|u_n|,
\end{equation}
for some fixed $\si \geq 0$. In what follows, we denote the space
of bounded linear operators from $l^2_{\sigma}$ to $l^2_{\sigma'}$ by $B(\sigma,\sigma')$
and from $l^1$ to $l^{\infty}$ by $B(1,\infty)$.

Our work is motivated by recent advances in analysis of the discrete Schr\"{o}dinger
operators in one dimension. Spectral theory and dispersive estimates
in $B(\sigma,-\sigma)$ for $H$ with compact $V$ were considered in \cite{KKK}
by extending the previous work on continuous wave and Schr\"{o}dinger equations
(these results were extended in \cite{KKV} to two-dimensional discrete operators).
Independently to this work, dispersive estimates
in $B(1,\infty)$ for $H$ with $V = 0$ were obtained in \cite{SK} by analyzing
integrals with fast oscillations. We shall extend the results of \cite{KKK}
and \cite{SK} to general potentials $V$ under some decay conditions at infinity.

Our ultimate goal in this work is to prove asymptotic stability of single-humped
solitons in the discrete one-dimensional nonlinear Schr\"{o}dinger equations.
Asymptotic stability of solitary waves in
continuous nonlinear Schr\"{o}dinger equations was considered in recent works
of Cuccagna \cite{cuccagna}, Perelman \cite{perelman}, and Schlag \cite{schlag}.
Although orbital stability of single-humped solitons in the discrete nonlinear
Schr\"{o}dinger equations has been proved long ago \cite{weinstein},
no work has been reported towards the proof of asymptotic stability
of single-humped solitons in the long-time evolution. This paper is the first step
in this direction, since linearization of the discrete nonlinear Schr\"{o}dinger equation
at a stationary soliton results in two Schr\"{o}dinger operators
$H$ with exponentially decaying potentials $V$ coupled together in a
non-self-adjoint operator \cite{PKF}. Results obtained in this work are expected to
be good for analysis of the non-self-adjoint matrix Schr\"{o}dinger operator, but
this will be the subject of the forthcoming work.

Let $R(\lambda) = \left( - \Delta + V - \lambda \right)^{-1}$ denote
the resolvent operator for $H = -\Delta + V$ and $R_0(\lambda) = (-\De - \lambda)^{-1}$
denote the free resolvent for $H_0 = -\Delta$. Since the
spectrum of $H_0$ is purely continuous and located on $[0,4]$,
we are particularly interested in the behavior of the resolvent
$R(\lambda)$ near the interval $[0,4]$ on ${\rm Im}\lambda = 0$.
We will use letter $\omega$ to indicate values of
$\lambda$ on the open interval $(0,4)$ and letter $\lambda$ to indicate values
on $\mathbb{C} \backslash [0,4]$.

Our article is structured as follows. We review properties of the
free resolvent $R_0(\lambda)$ in Section 2. These properties are used
to prove the limiting absorption principle for the resolvent $R(\lambda)$ on
$(0,4)$ in Section 3, the Puiseux expansions of the resolvent $R(\lambda)$ associated with
a generic potential $V$ near the end points $0$ and $4$ in Section
4, and the dispersive estimates on the time evolution of $\dot{u} = H u$
in Section 5. Appendices A, B, C, and D give proofs to technical lemmas used
in the main part of the article.

{\bf Acknowledgements.}
The authors thank P. Kevrekidis for stimulating collaborations. D.P.
is supported by NSERC. A.S. is supported in part by NSF-DMS, 0701802.

\section{Properties of the free resolvent}

Let $\lambda \in \mathbb{C} \backslash [0,4]$ and define $\theta =
\theta(\lambda)$ to be a unique solution of the transcendental equation
\begin{equation}
\label{trans-equation} 2 - 2 \cos \theta = \lambda
\end{equation}
in the domain $D=\{-\pi\leq {\rm Re} \theta \leq \pi, \; {\rm Im}
\theta<0\}$. If $\lambda = \omega \pm i \ve$ and $\ve
\downarrow 0$, then $\theta(\lambda) = \theta_{\pm}(\omega,\ve) + i
\nu_{\pm}(\omega,\ve)$ with
$$
2 - 2 \cos \theta_{\pm} = \omega + {\rm O}(\ve^2), \quad \nu_{\pm} =
\pm \frac{\ve}{2 \sin \theta_{\pm}} + {\rm O}(\ve^3).
$$
Since $\nu_{\pm} < 0$ for $\ve > 0$, we obtain that roots of $2 - 2
\cos \theta_{\pm} = \omega$ for $\omega \in (0,4)$ and $\ve = 0$ lie in
the intervals $\theta_+ \in (-\pi,0)$ and $\theta_- \in (0,\pi)$ with the symmetry
$\theta_+ = - \theta_-$.

Let $\phi$ solve the difference equation $(-\Delta - \lambda) \phi = f$
for any $f \in l^2(\mathbb{Z})$ and define the free resolvent operator $R_0(\lambda)$
by its solution $\phi = R_0(\lambda) f$. Then, direct substitution shows that
$R_0(\lambda)$ is explicitly represented by
\begin{equation}
\label{eq:1} \phi_n = (R_0(\lambda) f)_n = -\f{i}{2 \sin \theta} \suml_{m
\in \mathbb{Z}} e^{-i \theta |n-m|} f_m.
\end{equation}
In what follows, we summarize properties of the free resolvent. See Sections 2 and 3 in \cite{KKK}
for further details.

\begin{enumerate}
\item Since the sequence $\{ e^{-i \theta |n|} \}_{n \in \mathbb{Z}}$ is
exponentially decaying as $|n| \to \infty$ if ${\rm Im} \theta < 0$
and $l^2_{\sigma}$ is closed with respect to convolution for any $\sigma >
\frac{1}{2}$, we can see that $R_0(\lambda)$ is defined in $B(\sigma,\sigma)$
for any $\lambda \in \mathbb{C} \backslash [0,4]$ and
$\sigma > \frac{1}{2}$.

\item Since the sum of the double-infinite sequence
$$
\left\{ \frac{e^{-i \theta_{\pm}
|n-m|}}{(1+n^2)^{\sigma} (1+m^2)^\si} \right\}_{m,n \in \mathbb{Z}}
$$
on $m,n \in \mathbb{Z}$ is bounded if $\sigma > \frac{1}{2}$ and $|\sin
\theta_{\pm}| > 0$ if $\omega \in (0,4)$, then $R_0^{\pm}(\omega) =
\lim_{\ve \downarrow 0} R_0(\omega \pm i \ve)$ are Hilbert--Schmidt
operators in $B(\si,-\si)$ for $\omega \in (0,4)$ and $\sigma >
\frac{1}{2}$.

\item For any $\omega \in (0,4)$, the operators $R_0^{\pm}(\omega)$ maps
$l^1(\mathbb{Z})$ to $l^{\infty}(\mathbb{Z})$ since
$$
\| R_0^{\pm}(\omega) f\|_{l^{\infty}} \leq \frac{1}{2 |\sin \theta_{\pm}|} \| f \|_{l^1}
$$
and $|\sin \theta_{\pm}| > 0$ for $\omega \in (0,4)$.

\item The free resolvent $R_0^{\pm}(\omega)$ diverges near
$\omega = 0$ and $\omega = 4$ because $\sin \theta_{\pm}$ vanishes
in the limit. Without loss of generality, we consider only the
case $\omega = 0$, where $\theta_+ = \theta_- = 0$. Using the
asymptotic expansion
$$
\sin \theta(\lambda) = -\sqrt{\lambda - \frac{\lambda^2}{4}} =
-\sqrt{\lambda} \left( 1 + {\rm O}(\lambda) \right),
$$
where the minus sign is chosen to ensure that ${\rm Im} \theta <
0$ if $0 < \arg(\lambda) \leq \pi$ near $\lambda = 0$, we write a formal
Puiseux expansion of the free resolvent in the form
\begin{equation}
\label{Puiseux-free} (R_0(\lambda) f)_n = \f{i}{2\sqrt{\lambda}}
\suml_{m \in \mathbb{Z}} f_m - \frac{1}{2} \sum_{m \in \mathbb{Z}}
|n - m| f_m + r_0(\lambda) f,
\end{equation}
where $r_0(\lambda)$ is the remainder term and $0 < \arg(\lambda) \leq \pi$.
Substituting $\lambda = \omega$ for $R_0^+(\omega)$ and
$\lambda = \omega e^{2 \pi i}$ for $R_0^-(\omega)$, we obtain
Therefore,
\begin{equation}
\label{Puiseux-free-omega} (R_0^{\pm}(\omega) f)_n = \pm \f{i}{2\sqrt{\omega}}
\suml_{m \in \mathbb{Z}} f_m - \frac{1}{2} \sum_{m \in \mathbb{Z}}
|n - m| f_m + r_0^{\pm}(\omega) f,
\end{equation}
where $\omega > 0$ is small. The first two terms
in (\ref{Puiseux-free-omega}) are Hilbert--Schmidt operators in space
$B(\sigma,-\sigma)$ for $\sigma > \frac{3}{2}$, while the
remainder term $r_0^{\pm}(\omega)$ is estimated to be of the order of ${\rm
O}(\sqrt{\omega})$ in space $B(\sigma,-\sigma)$ for $\sigma >
\frac{5}{2}$. Therefore, if we fix $\sigma > \frac{5}{2}$, we can write
(\ref{Puiseux-free-omega}) in the form
\begin{equation}
\label{resolvent-expansion-free}
R_0^{\pm}(\omega) = \pm \frac{i R_{-1}}{\sqrt{\omega}} + R_0 + {\rm
O}(\sqrt{\omega}),
\end{equation}
where
$$
\left( R_{-1} f \right)_n = \frac{1}{2} (1,f), \qquad \left( R_0 f \right)_n =
-\frac{1}{2} \sum_{m \in \mathbb{Z}} |n-m| f_m
$$
and $\omega > 0$ is small.

\item Due to the symmetry $\theta(\lambda) = - \theta(\bar{\lambda})$ of roots of equation
(\ref{trans-equation}) for all $\lambda \in \mathbb{C} \backslash
[0,4]$, the following symmetry holds $R_0^-(\omega) = \bar{R}_0^+(\omega)$
for all $\omega \in (0,4)$. Therefore, it is sufficient to consider only $R_0^+(\omega)$
and drop the superscript from the rest of the article.
\end{enumerate}

\section{Limiting absorption principle}

To study how the resolvent operator $R(\lambda)$, defined for
$\lambda \in \mathbb{C} \backslash [0,4]$, is extended to the
interval $\omega \in (0,4)$, we shall use the standard resolvent properties
\begin{equation}
\label{resolvent-identity} R(\lambda) = ( I + R_0(\lambda) V)^{-1}
R_0(\lambda) = R_0(\lambda) (I + V R_0(\lambda))^{-1}.
\end{equation}
The second identity is due to the fact that the operators
$R_0(\lambda)$ and $R(\lambda)$ are self-adjoint in $l^2$ for
$\lambda \in \mathbb{C} \backslash [0,4]$. If
\begin{equation}
\label{potential-decay-condition}
\sup_{n \in \mathbb{Z}} (1 + n^2)^{\sigma} |V_n| < \infty
\end{equation}
for any fixed $\sigma > \frac{1}{2}$, then $V : l^2_{-\si} \mapsto
l^2_{\si}$. We note that if $V \in l^1_{2 \sigma-1}$, then the condition (\ref{potential-decay-condition})
is satisfied. Since $R_0(\omega) : l^2_{\si} \to l^2_{-\si}$ for
every fixed $\omega \in (0,4)$ and $\si > \frac{1}{2}$, then $V
R_0(\omega)$ is a bounded Hilbert--Schmidt operator in
$B(\sigma,\sigma)$ for $\sigma > \frac{1}{2}$. Therefore, the
operator $I + V R_0(\omega)$ is invertible in $l^2_\si$ if and
only if it has a trivial kernel. We will show that the kernel of $I
+ V R_0(\omega)$ is indeed trivial for $\omega \in (0,4)$,
which leads to the limiting absorption principle
formulated as follows.

\begin{theorem}
\label{theo:1} Fix $\sigma > \frac{1}{2}$ and
assume that $V \in l^1_{2\sigma - 1}$.
The resolvent $R(\lambda)=(-\De + V - \lambda)^{-1}$, defined
for $\lambda \in \mathbb{C} \setminus [0,4]$ as a bounded operator
on  $l^2$, satisfies
\begin{equation}
\label{eq:1011} \sup_{\ve \downarrow 0} \|R(\omega \pm i
\ve)\|_{B(\si,-\si)} < \infty.
\end{equation}
for any fixed $\omega \in (0,4)$. As a
consequence, there exist $R^{\pm}(\omega) = \lim_{\ve\downarrow 0}
R(\omega \pm i \ve)$ in the norm of $B(\si,-\si)$.
\end{theorem}

\begin{proof}
By property 5 and identity (\ref{resolvent-identity}),
the symmetry $R^-(\omega) = \bar{R}^+(\omega)$ holds so it
is sufficient to consider $R^+(\omega)$ and drop the superscript
from the formalism. We will show that $[I + V R_0(\omega)]^{-1} \in B(\si,\si)$
for any fixed $\omega \in (0,4)$ and $\si > \frac{1}{2}$. Since $R_0(\omega) \in B(\si,-\si)$
by property 2, the proof
of the theorem will follow from the second resolvent identity
(\ref{resolvent-identity}). To show that the kernel of $I + V
R_0(\omega)$ is trivial in $l^2_{\sigma}$ for any fixed
$\omega \in (0,4)$ and $\sigma > \frac{1}{2}$, we will assume the
opposite and obtain a contradiction.

Let $f \in l^2_{\si}$ be an eigenvector of $I + V R_0(\omega)$.
Then, it solves the difference equation
\begin{equation}
\label{eq:02} f_n - i V_n \suml_{m \in \mathbb{Z}}
\f{e^{-i \theta|n-m|}}{2\sin \theta} f_m=0, \qquad n \in \mathbb{Z}.
\end{equation}
Multiplying both sides of (\ref{eq:02}) by $\overline{f}_n / V_n$,
taking the imaginary part, and summing over $n \in \mathbb{Z}$, we
obtain
$$
{\rm Im} \left[ i \suml_{m,n\in \mathbb{Z}} e^{-i \theta |m-n|}
f_m \overline{f}_n \right] = \suml_{m,n \in \mathbb{Z}}
\cos(\theta (m-n) ) f_m\overline{f}_n =0,
$$
whence
\begin{equation}
\label{eq:01} \left|\suml_{m \in \mathbb{Z} } \cos(\theta m) f_m
\right|^2 + \left|\suml_{m \in \mathbb{Z}} \sin(\theta m ) f_m
\right|^2=0.
\end{equation}
Therefore, the eigenvector $f$ lies in the constrained subspace of
$l^2_{\si}$ of codimension two:
\begin{equation}
\label{constrained-space} \tilde{l}^2_\si = \left\{ f \in l^2_\si:
\quad \sum_{n \in \mathbb{Z}} \cos(\theta n) f_n =
\sum_{n \in \mathbb{Z}} \sin(\theta n) f_n = 0
\right\}.
\end{equation}
Define an operator $\tilde{R}_0 : l^2_\si\to l^2_{-\si}$ by
$$
(\tilde{R}_0 f)_n = -\suml_{m \in \mathbb{Z}}
\f{\sin(\theta |n-m|)}{2\sin \theta}
f_m.
$$
Then, $f + V \tilde{R}_0 f = 0$, which implies that $f$ may be taken
to be real-valued, which we assume henceforth. To restrict operator
$\tilde{R}_0$ to the subspace $\tilde{l}^2_\si$, we introduce
$(a_1,a_2)$ as solutions of the algebraic system
$$
\left|\begin{array}{l}
a_1 \cos \theta + a_2 \cos(2 \theta)=0 \\
a_1 \sin \theta + a_2 \sin(2 \theta)=1,
\end{array}\right.
$$
which is nonsingular, since its determinant equals to $\sin \theta
< 0$ for any fixed $\theta \in (-\pi,0)$. Let $\tilde{K}$ be an operator defined by
\begin{eqnarray*}
(\tilde{K} f)_n & = & V_n \tilde{R}_0 f_n - \left(\sum_m V_m
(\tilde{R}_0 f)_m \cos(\theta m) \right) \delta_{n,0}
\\ & \phantom{t} & \phantom{text} -  \left(\sum_m V_m
(\tilde{R}_0 f)_m \sin(\theta m)  \right) (a_1 \delta_{n,1} + a_2
\delta_{n,2}).
\end{eqnarray*}
Since $V \tilde{R}_0$ is a Hilbert--Schmidt operator in $B(\si,\si)$ under the condition 
(\ref{potential-decay-condition}),
then $\tilde{K}$ is a compact operator from
$\tilde{l}^2_{\si}$ to $\tilde{l}^2_{\si}$. If $f$ is an eigenvector of $f + V
\tilde{R}_0 f = 0$ and $f \in \tilde{l}^2_{\si}$, then $f_n +
V_n(\tilde{R}_0 f)_n =0$ and thus
\begin{eqnarray*}
& &  \sum_m V_m (\tilde{R}_0 f)_m \cos(\theta m)=-\sum_m f_m \cos(\theta m)=0,\\
& & \sum_m V_m (\tilde{R}_0 f)_m \sin(\theta m)=-\sum_m f_m
\sin(\theta m) = 0,
\end{eqnarray*}
such that $f_n + (\tilde{K} f)_n = f_n + V_n (\tilde{R}_0 f)_n=0$.
Therefore, if $f$ exists, then $-1\in
spec_{\tilde{l}^{2}_\si}(\tilde{K})$.

Now, we shall approximate the potential $V$ by the compactly
supported potential $V^{N}$ with the entries $V^N_n =
\sum_{j=-N+1}^{N-1}  V_j \delta_{n,j}$.  Let $\tilde{K}^{N}$  be the
compact operator obtained from the operator $\tilde{K}$ when $V$ is
replaced by $V^{N}$. If $-1\in spec_{\tilde{l}^{2}_\si}(\tilde{K})$,
then, by Lemma \ref{le:ap1} in Appendix A there exists a subsequence of eigenvalues $-(a_{N_j}+i
b_{N_j})$ of the operators $\tilde{K}^{N_j}$ with eigenvectors $f^{N_j} \in
\tilde{l}^2_\si$, so that $\lim_{j \to \infty} (a_{N_j} + i b_{N_j}) = 1$ and
$\lim_{j \to \infty} \|f^{N_j} - f \|_{\tilde{l}^2_\si} = 0$. For simplicity,
we drop the subscript $j$ from $N_j$. More
precisely, the eigenvectors satisfy
\begin{eqnarray}
\nonumber (a_N+ i b_N) f^N_n & + &  V^{N}_n (\tilde{R}_0 f^N)_n =
\left(\sum_m V^N_m (\tilde{R}_0 f^N)_m \cos(\theta_+ m) \right)
\delta_{n,0} \\ & \phantom{t} & + \left(\sum_m V_m^N (\tilde{R}_0
f^N)_m \sin(\theta_+ m)  \right) (a_1 \delta_{n,1} + a_2
\delta_{n,2}). \label{eq:090}
\end{eqnarray}
Equation (\ref{eq:090}) implies that the support of $f^N$ is finite.

Define the discrete Fourier transform $\cf:l^2(\mathbb{Z}) \to
L^2[0,2\pi]$ by
$$
\{u_n\}\longleftrightarrow \hat{u}(\xi) = {\cal F}(u) =
\suml_{n=-\infty}^\infty u_n e^{i n \xi}.
$$
Since $f^N$ has a compact support, then $\hat{f}^N(\xi) = \cf(f^N)$
is a trigonometric polynomial. Since $f^N$ belongs to $\tilde{l}^2_{\si}$,
it satisfies the two constraints in (\ref{constrained-space}), which implies that
$\hat{f}^N(\theta) = \hat{f}^N(-\theta) = 0$. Define a sequence $\psi^N$ via the
inverse Fourier transform of
$$
\hat{\psi}^N(\xi) = \frac{\hat{f}^N(\xi)}{2 - 2\cos \xi -\omega}.
$$
Since the denominator is equal to zero exactly at $\xi =
\pm \theta$, which are also among zeros of the numerator, and
since $\hat{f}^N(\xi)$ is a trigonometric polynomial in $\xi$,
we conclude that $\hat{\psi}^N(\xi)$ is a trigonometric
polynomial as well. Therefore, $\psi^N$ has a compact support. By
definition, $\psi^N$ is found from equation $(-\De - \omega) \psi^N
= f^N$, which is equivalent to the equation
\begin{eqnarray}
 \label{psi-equation} -\psi^N_{n+1} - \psi^N_{n-1} + (2-\omega) \psi^N_n +
\frac{V_n}{a_N + i b_N} \psi^N_n = \alpha_1 \delta_{n,0} + \alpha_2
(a_1 \delta_{n,1} + a_2 \delta_{n,2}),
\end{eqnarray}
where
\begin{eqnarray*}
\alpha_1 & = & \frac{1}{a_N + i b_N} \sum_m V^N_m
(\tilde{R}_0 f^N)_m \cos(\theta m) \\
\alpha_2 & = & \frac{1}{a_N + i b_N} \sum_m V_m^N (\tilde{R}_0
f^N)_m \sin(\theta m).
\end{eqnarray*}
The only compact support solution of \eqref{psi-equation} has a
non-zero value at $n = 1$. Therefore, the eigenvector $f^N$ has a
compact support at $n = \{ 0,1,2\}$. By $\lim_{N \to \infty} \|f^N -
f \|_{\tilde{l}^2_\si} = 0$, we conclude that the support of $f$ is
also at the three nodes $n = \{ 0,1,2\}$. Therefore, the function $\psi$,
defined by a solution of $(-\De - \omega)\psi = f$, is also
compactly supported at $n = 1$. However, $\psi$ is also a solution of $(-\De +
V - \omega) \psi=0$ and the only compact support solution of this
equation is  $\psi\equiv 0$. Hence $f\equiv 0$, and we obtain a
contradiction. This contradiction implies, of course, that $I + V
R_0(\omega)$ is an invertible operator on $l^2_\si$, as claimed.
\end{proof}

\section{Puiseux expansions at the spectral edges}

The free resolvent $R_0(\omega)$ has a singular behavior as
$\omega \downarrow 0$, as follows from the expansion
(\ref{Puiseux-free-omega}). Recall that the superscripts are
dropped for $R_0(\omega)$ and $R(\omega)$. We will show that the resolvent operator
$R(\omega)$ has a regular behavior in the same limit provided
that $V$ is a generic potential in the following sense.

\begin{definition}
\label{definition-generic-potential}
$V \in l^1_1$ is called a generic potential if no solution
$\psi_0$ of equation $(-\Delta + V) \psi_0 = 0$ exists in
$l^2_{-\sigma}$ for $\frac{1}{2} < \sigma \leq \frac{3}{2}$.
\end{definition}

\begin{remark}
\label{remark-potential}
We show in Appendix B that solutions of $(-\Delta + V) \psi_0 = 0$
always belong to $l^2_{-\sigma}$ for $\sigma > \frac{3}{2}$.
\end{remark}

Since $R_0(\omega) : l^2_{\si} \mapsto l^2_{-\si}$ for any fixed
$\omega \in (0,4)$ and $\sigma > \frac{1}{2}$ and since $V : l^2_{-\si} \mapsto l^2_{\si}$
under the condition (\ref{potential-decay-condition}),
$T(\omega) = I + R_0(\omega) V$ is a bounded Hilbert--Schmidt operator
in $B(-\si,-\si)$ for $\sigma > \frac{1}{2}$. Since $R_0(\omega)$ is
represented by the expansion (\ref{resolvent-expansion-free}) for $\sigma
> \frac{5}{2}$, we obtain
$$
T(\omega) = \frac{i T_{-1}}{\sqrt{\omega}} + T_0 + {\rm
O}(\sqrt{\omega}),
$$
where
$$
(T_{-1} f)_n = \frac{1}{2} \sum_{m \in \mathbb{Z}} V_m f_m, \quad
(T_0 f)_n = f_n - \frac{1}{2} \sum_{m \in \mathbb{Z}} |n-m| V_m f_m
$$
and $\omega > 0$ is small.
We shall denote $\tilde{T}(\omega) = T(\omega) - \frac{i
T_{-1}}{\sqrt{\omega}} = T_0 + {\rm O}(\sqrt{\omega})$.
We will show that the operator
$\tilde{T}(\omega)$ is invertible near $\omega = 0$ if
no solution $u_0$ of equation $T_0 u_0
= 0$ exists in $l^2_{-\si}$ for any $\sigma > \frac{3}{2}$ and
$\langle V, T_0^{-1} 1 \rangle \neq 0$, where angular brackets
denote inner products in $l^2$ and $1$ is the vector with $1_n = 1$
for all $n \in \mathbb{Z}$. Lemma \ref{lemma-equivalence} in Appendix B
shows that this condition is equivalent to the condition that $V$
is a generic potential of Definition \ref{definition-generic-potential}.
Puiseux expansion of the resolvent near $\omega = 0$ is defined in
the following theorem.

\begin{theorem}
\label{theo:2} Fix $\sigma > \frac{5}{2}$ and
assume that $V \in l^1_{2\si-1}$ is generic in the sense: no solution $u_0$ of
equation $T_0 u_0 = 0$ exists in $l^2_{-\si}$ for any $\sigma >
\frac{3}{2}$ and $\langle V, T_0^{-1} 1 \rangle \neq 0$. The
resolvent $R(\omega)$, defined for $\omega \in (0,4)$, has the
expansion
\begin{equation}
\label{eq:Puiseux} R(\omega) = R(0) + {\rm O}(\sqrt{\omega})
\end{equation}
in the norm of $B(\sigma,-\sigma)$ for sufficiently small $\omega >
0$.
\end{theorem}

\begin{proof}
We will show first that $\tilde{T}(\omega) \in B(-\si,-\si)$ is
invertible for any fixed small $\omega > 0$ and any fixed $\si >
\frac{3}{2}$, provided that the potential $V \in l^1_{2\si-1}$ is generic. Let $u =
u^{\perp} + c(\omega) V$, where $c(\omega)$ is a coefficient
and $u^{\perp}$ satisfies the orthogonal projection
$\langle V, u^{\perp} \rangle = 0$. For any $f \in l^2_{-\si}$,
equation $T(\omega) u = f$ is equivalent to
\begin{equation}
\label{eq:LS} \tilde{T}(\omega) u^{\perp} + c(\omega) T(\omega) V =
f.
\end{equation}
Since $V$ is decaying as $|n| \to \infty$, we have
$u^{\perp} \in l^2_{-\si}$ if and only if $u \in l^2_{-\sigma}$
for any fixed $\si > \frac{1}{2}$.
Since $\tilde{T}(\omega) - T_0 \to 0$ as $\omega \downarrow 0$ in
$B(-\sigma,-\sigma)$ for $\sigma > \frac{3}{2}$, the operator
$\tilde{T}(\omega)$ is invertible if $T_0$ is invertible.
Under the condition (\ref{potential-decay-condition}), we have
$$
\sum_{m,n \in \mathbb{Z}} \frac{|n-m|^2 V_m^2
(1+m^2)^{\sigma}}{(1+n^2)^{\sigma}} \leq C \sum_{m,n \in \mathbb{Z}}
\frac{|n-m|^2}{(1+m^2)^{\sigma}(1+n^2)^{\sigma}} < \infty
$$
for some $C > 0$ and any $\sigma > \frac{1}{2}$. Therefore,
$\tilde{T}_0 = T_0 - I$ is a
Hilbert--Schmidt operator in $B(-\si,-\si)$ for any fixed
$\si > \frac{3}{2}$, such that $T_0$ is invertible if and
only if the kernel of $T_0$ is empty in $l^2_{\si}$ for $\si > \frac{3}{2}$,
which is a condition that $V$ is a generic potential.

Since $\tilde{T}(\omega)$ is invertible for sufficiently small
$\omega > 0$, a unique solution of (\ref{eq:LS}) is
$$
u^{\perp} = [\tilde{T}(\omega)]^{-1} (f - c(\omega) T(\omega) V) = -
c V + [\tilde{T}(\omega)]^{-1} \left( f -
\frac{i c(\omega)}{\sqrt{\omega}}  T_{-1} V\right).
$$
To find uniquely the coefficient $c(\omega)$ in the decomposition $u
= u^{\perp} + c(\omega) V$, we let $S(\omega) =
[\tilde{T}(\omega)]^{-1}$ and define the adjoint operators
$[S(\omega)]^*$ and $[\tilde{T}(\omega)]^*$ as bounded maps in $B(\si,\si)$
for any fixed $\sigma > \frac{3}{2}$. Since $\| V \|_{l^2_s} \leq \| V \|_{l^1_s}$,
if $V \in l^1_s$ for any fixed $s \geq 2 \si - 1$, then $V \in l^2_{\si}$,
such that $W = S^* V \in l^2_{\si}$ for any $\sigma > \frac{3}{2}$.

Let us now fix $\sigma > \frac{5}{2}$ and represent
$\tilde{T}(\omega)$ by $T_0 + {\rm O}(\sqrt{\omega})$. Then,
$S(\omega) = S_0 + {\rm O}(\sqrt{\omega})$, where $S_0 = T_0^{-1}$.
Using the inner products in $l^2$, we obtain $\langle W,
\tilde{T}(\omega) u^{\perp} \rangle = \langle V, u^{\perp} \rangle =
0$, such that
$$
c(\omega) = \frac{\langle W, f \rangle}{\langle W, T(\omega) V
\rangle} = \frac{\langle V, S(\omega) f \rangle}{\langle V,
S(\omega) T(\omega) V \rangle},
$$
provided that $\langle V, S(\omega) T(\omega) V \rangle = \| V
\|^2_{l^2} \left( 1 + \frac{i}{2 \sqrt{\omega}} \langle V, S(\omega)
1 \rangle \right) \neq 0$ for sufficiently small $\omega > 0$. Since
$S(\omega) - S_0 \to 0$ as $\omega \downarrow 0$ in $B(\si,\si)$ for
$\sigma > \frac{3}{2}$, this condition is satisfied if $\langle V,
S_0 1 \rangle = \langle V, T_0^{-1} 1 \rangle \neq 0$, which is true
for generic potentials $V$.

The first resolvent identity (\ref{resolvent-identity}) implies that
if $T(\omega) u = f$ for some $f \in l^2_{-\si}$ and $f =
R_0(\omega) \psi$ for some $\psi \in l^2_{\si}$, then $u =
R(\omega) \psi$ for a fixed $\omega \in (0,4)$. We shall now
finish the proof of theorem by computing the limit $\omega
\downarrow 0$ in the following chain of identities:
\begin{equation}
\label{technical-equation} R(\omega) \psi = u = u^{\perp} +
c(\omega) V = [\tilde{T}(\omega)]^{-1} \left( R_0^+(\omega) \psi -
\frac{i c(\omega)}{\sqrt{\omega}} T_{-1} V \right),
\end{equation}
where
$$
c(\omega) = \frac{\langle V, S(\omega) R_0(\omega) \psi
\rangle}{\langle V, S(\omega) T(\omega) V \rangle} =
\frac{\frac{i}{2\sqrt{\omega}} \langle V, S(\omega) 1 \rangle
\langle 1, \psi \rangle +  \langle V, S(\omega) \tilde{R}_0(\omega)
\psi \rangle}{\| V \|^2_{l^2} \left( \frac{i}{2\sqrt{\omega}}
\langle V,  S(\omega) 1 \rangle + 1 \right)},
$$
where $\tilde{R}_0(\omega) = R_0(\omega) - \frac{i
R_{-1}}{\sqrt{\omega}}$. Therefore, $\lim_{\omega \downarrow 0}
c(\omega) = c(0)$ exists and the singular term of
(\ref{technical-equation}) is canceled because of the explicit
computation
$$
\frac{i}{2 \sqrt{\omega}} \sum_{m \in \mathbb{Z}} \psi_m -
\frac{i c(0)}{\sqrt{\omega}} T_{-1} V = \frac{i}{2 \sqrt{\omega}}
\left[ \langle 1, \psi \rangle - \frac{\langle 1, \psi \rangle \|
V\|^2_{l^2}}{\| V \|^2_{l^2}} \right] = 0.
$$
As a result, the expansion (\ref{eq:Puiseux}) is proved with
$$
R(0) \psi = S_0 R_0 \psi + \left(\frac{\langle 1, \psi
\rangle}{\langle V, S_0 1 \rangle} - \frac{\langle V, S_0 R_0 \psi
\rangle}{\langle V, S_0 1 \rangle} \right)  S_0 1,
$$
where we have used again that $\langle V, S_0 1 \rangle \neq 0$ for
generic potentials.
\end{proof}

\begin{remark}
Not only Theorem \ref{theo:2} generalizes Theorems 5.1 and 6.1 in
\cite{KKK} from compact to spatially decaying potentials $V$ but
also the class of generic potentials $V$ is defined more precisely
compared to Definition 5.1 in \cite{KKK}. In addition, the values of
$\sigma$ can be taken for $\sigma > \frac{5}{2}$ compared to $\sigma
> \frac{7}{2}$ in \cite{KKK}, since no terms of ${\rm O}(\sqrt{\omega})$ in the expansions of
$T(\omega)$ and $S(\omega)$ are used to obtain the leading order
term of $R(0)$.
\end{remark}

\section{Dispersive estimates}

Using the previous analysis of the resolvent operator $R(\omega)$,
we switch our focus to the discussion of the dispersive estimates for the
time-dependent discrete Schr\"odinger equation $u_t = H u$ with initial data
$u(0) = u_0$ in an appropriate function space. We have two types of dispersive
estimates. The first one describes decay of the
semigroup $e^{i t H}$ acting on the weighted $l^2$ spaces and it is an extension of
Theorem 7.1 in \cite{KKK}. The second, more delicate estimate describes decay of
the semigroup that maps $l^1$ into $l^{\infty}$ and it
is an extension of the dispersive estimate of the free resolvent in \cite{SK}.

Let $P_j$ denote projections on the eigenspaces corresponding to the eigenvalues
$\omega_j \in \mathbb{R} \backslash [0,4]$ of the self-adjoint operator $H$. We shall prove that
the discrete spectrum is finite-dimensional, such that $j$ can be enumerated
from $j = 1$ to $j = n < \infty$. By the spectral theory, projection to the essential
(absolutely continuous) spectrum of $H$ is defined by $P_{\rm a.c.}(H) = I - \sum_{j=1}^n P_j$.

\begin{lemma}
\label{lemma-discrete-spectrum}
Fix $\sigma > \frac{5}{2}$ and assume that $V \in l^1_{2\si-1}$
is generic in the sense of Definition \ref{definition-generic-potential}.
The discrete spectrum of $H$ is finite-dimensional and located in the two
segments $(\omega_{\rm min},0) \cup (4,\omega_{\rm max})$, where
$$
\omega_{\rm min} = \min_{n \in \mathbb{Z}} \{ 0, V_n \}, \qquad
\omega_{\rm max} = \max_{n \in \mathbb{Z}} \{ 4, 4 + V_n \}
$$
\end{lemma}

\begin{proof}
Since $H$ is self-adjoint in $l^2(\mathbb{Z})$, eigenvalues of the discrete
spectrum are all located on $\mathbb{R}$. By Theorem \ref{theo:1}, no embedded
eigenvalues may occur in $(0,4)$ if $\sigma > \frac{1}{2}$.
By Theorem \ref{theo:2}, no eigenvalues are located at $0$ and $4$ if
$\sigma > \frac{5}{2}$ and $V$ is a generic potential in the sense of
Definition \ref{definition-generic-potential}.
The upper and lower bounds on the location of eigenvalues follows from the fact that
$0 \leq (\phi,-\Delta \phi) \leq 4 \| \phi \|^2_{l^2}$. Since the
resolvent operator $R(\lambda)$ is a meromorphic function
on $\lambda \in \mathbb{R}\backslash [0,4]$ with bounded limits
in $B(\si,-\si)$ for $\si > \frac{5}{2}$ as $\lambda \to 0$ and
$\lambda \to 4$, it has a finite number of poles in the compact domain
$(\omega_{\rm min},0) \cup (4,\omega_{\rm max})$. Therefore,
the discrete spectrum of $H$ is finite-dimensional.
\end{proof}

\begin{remark}
Unlike the continuous Schr\"{o}dinger operator, isolated eigenvalues of the
discrete Schr\"{o}dinger equation outside $[0,4]$ can be supported by the
potential $V$ with the range in $[0,4]$. Appendix in \cite{KKK} give examples of
such eigenvalues for compact potentials supported at one or two nodes
with any non-zero values of $V$.
\end{remark}

The results on the dispersive estimates for the one-dimensional discrete
Schr\"odinger equation are described in the following two theorems.

\begin{theorem}
\label{theo:dec1}
Fix $\sigma > \frac{5}{2}$ and assume that $V \in l^1_{2\si-1}$
is generic in the sense of Definition \ref{definition-generic-potential}.
Then, there exists a constant $C$ depending on $V$, so that
\begin{equation}
\label{eq:15}
\left\| e^{i t H}P_{a.c.}(H) \right\|_{l^2_\si\to l^2_{-\si}}\leq C t^{-3/2}
\end{equation}
for any $t > 0$.
\end{theorem}

\begin{proof}
The proof of \eqref{eq:15} is standard and it follows the outline in \cite{KKK}. By
the Cauchy formula in $B(\si,-\si)$, we obtain
\begin{eqnarray}
\label{eq:m1}
e^{i t H}P_{a.c.}(H) = \f{1}{2 \pi i} \int_0^4 e^{i t \om} \left[ R^+(\om)
- R^-(\om) \right] d\om = \f{1}{\pi} \int_0^4 e^{i t \om} {\rm Im} R(\om) d\om,
\end{eqnarray}
where we have dropped the superscript for $R^+(\om)$.
By the representation of the perturbed resolvent in Theorem \ref{theo:2} (in
particular, by the fact that $R(0)$ is real), we have
\begin{eqnarray}
\label{eq:76}
\left\| R(\om) \right\|_{B(\si,-\si)}
\leq C, \quad \left\| {\rm Im} R(\om) \right\|_{B(\si,-\si)}
\leq C \om^{1/2}, \end{eqnarray}
and
\begin{eqnarray}
\label{eq:77}
& & \left\|  \f{d^j}{d \omega^j} R(\om) \right\|_{B(\si,-\si)}
\leq C \om^{1/2-j},
\quad j=1,2
\end{eqnarray}
for some $C > 0$.
Introduce smooth cutoff functions $\chi_1, \chi_2\in C^\infty_0$, so
that $\chi_1 + \chi_2=1$ for all $\om\in [0,4]$, while
${\rm supp}(\chi_1) \subset [0,3]$ and ${\rm supp}(\chi_2) \subset [1,4]$. Write
\begin{equation}
\label{projection-representation}
e^{i t H}P_{a.c.}(H) = \f{1}{2\pi i} \int_0^3 e^{i t \om} \chi_1(\om) {\rm Im} R(\om) d\om+
\f{1}{2\pi i}\int_1^4 e^{i t \om} \chi_2(\om) {\rm Im} R(\om) d\om.
\end{equation}
To each of the two terms, one can apply the Lemma \ref{le:90} from Appendix C. Note that the conditions
on the function $F(w)= {\rm Im} R(\om)\chi_1(\om) \in B(\si, -\si)$ are satisfied,
because of the bounds \eqref{eq:76} and  \eqref{eq:77}.
\end{proof}

\begin{theorem}
\label{theo:dec2}
Fix $\sigma > \frac{5}{2}$ and assume that $V \in l^1_{2\si-1}$ is
generic in the sense of Definition \ref{definition-generic-potential}.
Then, there exists a constant $C$ depending on $V$, so that
\begin{equation}
\label{eq:16}
\left\| e^{i t H}P_{a.c.}(H) \right\|_{l^1\to l^\infty}\leq C t^{-1/3}
\end{equation}
for any $t > 0$.
\end{theorem}

To prove Theorem \ref{theo:dec2}, we develop scattering theory for fundamental solutions
of the discrete Schr\"{o}dinger equation, following the work \cite{GS}
for the continuous Schr\"{o}dinger equation.
Let $\psi^{\pm}$ be two linearly independent solutions of
\begin{equation}
\label{Schrodinger-scattering}
\psi_{n+1} + \psi_{n-1} + (\om - 2) \psi_n = V_n \psi_n,
\end{equation}
according to the boundary conditions $\left| \psi^{\pm}_n - e^{\mp i n \theta} \right| \to 0$
as $n \to \pm \infty$, where $\theta$ is a root of
\begin{equation}
\label{transformation-theta}
2 - 2 \cos \theta = \omega
\end{equation}
for $\omega \in [0,4]$. Since the solutions depend on $\theta$, we may
use $\psi^{\pm}(\theta)$ instead of $\psi^{\pm}$. The Green function
representation of the two solutions is
\begin{eqnarray*}
\psi^+_n(\theta) & = & e^{-i n \theta} - \frac{i}{2 \sin \theta}
\sum_{m = n}^{\infty} \left( e^{i \theta (m-n)} - e^{-i \theta (m-n)}
\right) V_m \psi^+_m(\theta),
\\ \psi^-_n(\theta) & = & e^{i n \theta} - \frac{i}{2 \sin \theta}
\sum_{m = -\infty}^{n} \left( e^{i \theta (n-m)} - e^{-i \theta (n-m)}
\right) V_m \psi^-_m(\theta).
\end{eqnarray*}
Let $\psi^{\pm}_n(\theta) = e^{\mp i n \theta} f_n^{\pm}(\theta)$
for all $n \in \mathbb{Z}$. Writing the Green function representation for $f^+(\theta)$
in the form
\begin{eqnarray}
\label{f-plus-theta}
f^+_n(\theta) = 1 - \frac{i}{2 \sin \theta}
\sum_{m = n}^{\infty} \left( 1 - e^{-2 i \theta (m-n)}
\right) V_m f^+_m(\theta)
\end{eqnarray}
and using the formula
$$
\left| \frac{1 - e^{-2i \theta x}}{2 i \sin \theta} \right| \leq C_0 |x|, \quad \forall \theta \in [-\theta_0,\theta_0], \;\; \forall x \in \mathbb{R},
$$
for some $C_0 > 0$ and fixed $0 < \theta_0 < \frac{\pi}{2}$, it follows
from the Neumann series that the sequence $\{ f^+_n(\theta) \}_{N_0}^{\infty}$ is
uniformly bounded in $l^{\infty}$ on $[-\theta_0,\theta_0]$
if $V \in l^1_1$, where $N_0$ is defined as the smallest integer, for which
$$
C_0 \sum_{k = 1}^{\infty} k |V_{N_0 + k}| < 1.
$$
Moreover, the sequence $\{ f^+_n(\theta) \}_{N_0}^{\infty}$ is
analytically continued in the strip $\Sigma_0 = \{ -\theta_0 \leq {\rm Re}\theta \leq \theta_0, \;\; {\rm Im} \theta \geq 0\}$, such that $\partial_{\theta} f^+(\theta)$ exists in the interior of $\Sigma_0$.
By taking the derivative of (\ref{f-plus-theta}) in $\theta$, we obtain
$$
\partial_{\theta} f^+_n(\theta) =
\sum_{m = n}^{\infty} \partial_{\theta} \left( \frac{1 - e^{-2 i \theta (m-n)}}{2 i \sin \theta}
\right) V_m f^+_m(\theta) - \frac{i}{2 \sin \theta}
\sum_{m = n}^{\infty} \left( 1 - e^{-2 i \theta (m-n)}
\right) V_m  \partial_{\theta} f^+_m(\theta).
$$
By the same argument, it follows
from the Neumann series that the sequence $\{ \partial_{\theta} f^+_n(\theta) \}_{N_0}^{\infty}$ is
uniformly bounded in $l^{\infty}$ on $[-\theta_0,\theta_0]$
if $V \in l^1_2$. Therefore, if $V \in l^1_2$, then
$$
\sup_{\theta \in [-\theta_0,\theta_0]} \left(
\| \partial_{\theta} f^+(\theta) \|_{l^{\infty}([N_0,\infty))} +
\| \partial_{\theta} f^+(\theta) \|_{l^{\infty}([N_0,\infty))} \right) < \infty.
$$
If $N_0 > 0$, then the above bound can be extended in $l^{\infty}(\mathbb{Z}_+)$
since the finite sequence $\{ f_n^+(\theta) \}_{n=0}^{N_0}$ satisfies a second-order difference equation
with analytic coefficients and analytic boundary values $f_{N_0}^+(\theta)$ and $f_{N_0+1}^+(\theta)$
in the strip $\Sigma_0$. Similar estimates hold for $\{ f_n^-(\theta) \}_{n\in \mathbb{Z}_-}$. Thus,
if $V \in l^1_2(\mathbb{Z})$, then
$\{ f_n^{\pm}(\theta) \}_{n \in \mathbb{Z}_{\pm}}$ are analytic in the strip $\Sigma_0$ and
there exist uniform bounds
\begin{equation}
\label{bounds-F-pm}
F_{\pm} = \sup_{\theta \in [-\theta_0,\theta_0]} \left(
\| f^{\pm}(\theta) \|_{l^{\infty}(\mathbb{Z}_{\pm})} +
\| \partial_{\theta} f^{\pm}(\theta) \|_{l^{\infty}(\mathbb{Z}_{\pm})} \right) < \infty.
\end{equation}
Let us define the discrete Wronskian
\begin{equation}
\label{Wronskian}
W[\psi^+,\psi^-] = \psi^+_n \psi^-_{n+1} - \psi^+_{n+1} \psi^-_n \equiv W(\theta),
\end{equation}
which is independent of $n \in \mathbb{Z}$ and analytic in $\Sigma_0$.
The discrete Green
function for the resolvent operators $R^{\pm}(\omega)$ has the kernel
$$
\left[ R^{\pm}(\omega) \right]_{n,m} = \frac{1}{W(\theta_{\pm})} \left\{ \begin{array}{cc}
\psi_n^+(\theta_{\pm}) \psi_m^-(\theta_{\pm}) \;\; \mbox{for} \;\; n \geq m \\
\psi_m^+(\theta_{\pm}) \psi_n^-(\theta_{\pm}) \;\; \mbox{for} \;\; n < m \end{array} \right.
$$
where $\theta_- = -\theta_+$ and
$\theta_- \in [0,\pi]$ for $\omega \in [0,4]$ (see Section 2).
Using (\ref{eq:m1}), we represent
$e^{i t H} P_{a.c.}(H)$ by its kernel for $n < m$:
\begin{eqnarray}
\nonumber
\left[ e^{i t H}P_{a.c.}(H) \right]_{n,m} & = & \f{1}{2 \pi i } \int_0^4 e^{i t \om}
\left[ \frac{\psi_m^+(\theta_+) \psi_n^-(\theta_+)}{W(\theta_+)}
- \frac{\psi_m^+(\theta_-) \psi_n^-(\theta_-)}{W(\theta_-)} \right] d\om \\
& = & \f{i}{\pi} \int_{-\pi}^{\pi} e^{i t (2 - 2\cos \theta)}
\frac{\psi_m^+(\theta) \psi_n^-(\theta)}{W(\theta)}  \sin \theta d \theta,
\label{eq:m11}
\end{eqnarray}
where we have unfolded the branch points $\omega = 0$ and $\omega = 4$ by using
the transformation (\ref{transformation-theta}). If
$V$ is a generic potential in the sense of
Definition \ref{definition-generic-potential},
then Appendix B shows that the two solutions $\psi^{\pm}(0)$ are
linearly independent, such that $W(0) \neq 0$ (the point
$\theta = 0$ corresponds to $\omega = 0$). A similar analysis
applies to the points $\theta = \pm \pi$ which correspond
to $\omega = 4$.

Let $\chi_0, \chi \in C^{\infty}_0:\chi_0(\theta)+\chi(\theta)=1$ for
all $\theta\in [-\pi, \pi]$, so that
${\rm supp} \chi_0 \subset [-\theta_0,\theta_0] \cup
(-\pi, -\pi+\theta_0) \cup (\pi-\theta_0, \pi)$ and
${\rm supp} \chi \subset [\theta_0/2,\pi-\theta_0/2]
\cup [-\pi+\theta_0/2,-\theta_0/2]$, where
 $0< \theta_0 \leq  \frac{\pi}{4}$. Here the value $\theta_0$ is the same number,
which is used in the bounds (\ref{bounds-F-pm}). (If the original number
$\theta_0>\pi/4$, we reassign it to be $\theta_0=\pi/4$.) It is important for
our argument that the support of $\chi$ stays away (by a fixed number
$\theta_0/2$!) from both $0$ and $\pi$.

We can now formulate and prove two technical lemmas
needed for the proof of Theorem \ref{theo:dec2}.

\begin{lemma}
\label{le:7}
Assume  $V \in l^1_2$ and $W(0) \neq 0$.  Then, there exists $C > 0$ such that
\begin{equation}
\label{eq:m21}
\sup_{n < m} \left| \int_{-\pi}^{\pi} e^{i t (2 - 2\cos \theta)} \chi_0(\theta)
\frac{\psi_m^+(\theta) \psi_n^-(\theta) \sin \theta}{W(\theta)} d \theta \right| \leq C t^{-1/2}
\end{equation}
for any $t > 0$.
\end{lemma}

\begin{proof}
The proof is different for regions $n < 0 < m$, $0 < n < m$ and $n < m < 0$.
In the case $n < 0 < m$, we write
\begin{eqnarray*}
& \phantom{t} &
\sup_{n < 0 < m} \left| \int_{-\pi}^{\pi}
e^{i t (2 - 2\cos \theta)} \chi_0(\theta)
\frac{\psi_m^+(\theta) \psi_n^-(\theta)}{W(\theta)}
\sin \theta d \theta \right| =\\
& & \sup_{n < 0 < m} \left| \int_{-\theta_0}^{\theta_0}
e^{i t (2 - 2\cos \theta)} e^{i(n-m) \theta} g_{n,m}(\theta) d \theta \right|,
\end{eqnarray*}
where $g_{n,m}(\theta) = \chi_0(\theta)
\frac{f_m^+(\theta) f_n^-(\theta)}{W(\theta)} \sin \theta$. Since $f_n^+(\theta)$
and $f_m^-(\theta)$ are continuously differentiable on $[-\theta_0,\theta_0]$ and
satisfy the uniform bounds (\ref{bounds-F-pm}) if $V \in l^1_2$ and
since $|W(\theta)| \geq W_0 > 0$ if $W(0) \neq 0$,
the dispersive estimate (\ref{eq:m21}) follows by Lemma \ref{le:91} of Appendix D.

In the case $0 < n < m$, the above
estimate is not sufficient since $f^-_n(\theta)$ grows linearly as $n \to \infty$. Therefore, we use
the scattering theory for fundamental solutions of (\ref{Schrodinger-scattering}) and represent
$$
\psi^-(\theta) = a(\theta) \psi^+(\theta) + b(\theta) \psi^+(-\theta),
$$
where
$$
a(\theta) = \frac{W[\psi^-(\theta),\psi^+(-\theta)]}{2 i \sin \theta}, \quad
b(\theta) = \frac{W[\psi^-(\theta),\psi^+(\theta)]}{-2 i \sin \theta} = \frac{W(\theta)}{2 i \sin \theta},
$$
and the discrete Wronskian is defined by (\ref{Wronskian}). As a result,
we write
\begin{eqnarray*}
& \phantom{t} &
\sup_{0 < n < m} \left| \int_{-\pi}^{\pi} e^{i t (2 - 2\cos \theta)} \chi_0(\theta)
\frac{\psi_m^+(\theta) \psi_n^-(\theta)}{W(\theta)}  \sin \theta d \theta \right| \\
& \leq & \frac{1}{2} \sup_{0 < n < m} \left| \int_{-\pi}^{\pi} e^{i t (2 - 2\cos \theta)} e^{i -(n + m) \theta}
\chi_0(\theta) \frac{f_m^+(\theta) f_n^+(\theta) W[\psi^-(\theta),\psi^+(-\theta)]}{W(\theta)}
d \theta \right|  \\
& \phantom{t} & + \frac{1}{2} \sup_{0 < n < m} \left|
\int_{-\pi}^{\pi} e^{i t (2 - 2\cos \theta)} e^{-i (m - n) \theta}
\chi_0(\theta) f_m^+(\theta) f_n^+(-\theta) d \theta \right|.
\end{eqnarray*}
Each term here is estimated by the bound \eqref{eq:780} of Lemma \ref{le:91} for
an appropriate function $g(\theta)$.
The last case $n < m < 0$ is estimated similarly to the
case $0 < n < m$ by using the scattering theory for $\psi^+(\theta)$
in terms of $\psi^-(\theta)$ and $\psi^-(-\theta)$.
\end{proof}

It remains to treat the case, when the cutoff $\chi$ is placed on
$[\theta_0/2,\pi-\theta_0/2]
\cup [-\pi+\theta_0/2,-\theta_0/2]$, where $0 < \theta_0 \leq \frac{\pi}{4}$
is a fixed number. Using the original representation (\ref{eq:m1}),
we need to estimate the operator norm of
$$
I_V = \int_{-\pi}^{\pi} e^{ i t (2-2\cos \theta)} \chi(\theta)
{\rm Im} R(2 - 2 \cos \theta) \sin \theta d\theta
$$
in $B(1,\infty)$. This estimate is given by the following lemma. Clearly,
Lemma \ref{le:7} and Lemma \ref{le:8} imply Theorem  \ref{theo:dec2}.

\begin{lemma}
\label{le:8}
Fix $\sigma > \frac{5}{2}$ and assume that $V \in l^1_s$ for $1 \leq s < 2\si-1$.
Then, there exists $C > 0$ such that
\begin{equation}
\label{eq:m22}
\left\| \int_{-\pi}^{\pi} \int_{-\pi}^{\pi} e^{ i t (2-2\cos \theta)} \chi(\theta)
{\rm Im} R(2 - 2 \cos \theta) \sin \theta d\theta \right\|_{B(1,\infty)} \leq C
t^{-1/3}
\end{equation}
for any $t > 0$.
\end{lemma}

\begin{proof}
We start by recalling the finite Born series
$$
R(\omega) = R_0(\omega) - R_0(\omega) V R_0(\omega) +
R_0(\omega) V R(\omega) V R_0(\omega),
$$
which follows by iterating the resolvent identity (\ref{resolvent-identity}).
We can write $I_V = I^1-I_V^2+I_V^3$, where
\begin{eqnarray*}
& & I^1 =  \int_{-\pi}^{\pi}  e^{ i t (2-2\cos \theta)}
\chi(\theta) {\rm Im} R_0(2-2\cos \theta) \sin\theta d\theta \\
& & I_V^2 = \int_{-\pi}^{\pi}  e^{ i t (2-2\cos \theta)}
\chi(\theta) {\rm Im} R_0(2-2\cos \theta) V R_0(2-2\cos \theta) \sin\theta d\theta \\
& & I_V^3 = \int_{-\pi}^{\pi}  e^{ i t (2-2\cos \theta)}
\chi(\theta) {\rm Im} R_0(2-2\cos \theta) G_V(\theta) R_0(2-2\cos \theta) \sin\theta d\theta
\end{eqnarray*}
where $G_V(\theta):= V R(2-2\cos \theta) V$. For $I^1$,
we observe that this is in fact a solution of the free Schr\"odinger
equation and can be written as
$$
(I^1 f)_n = -\frac{1}{2} \sum_{m \in \mathbb{Z}} f_m
\int_{-\pi}^{\pi}  e^{i t (2- 2\cos \theta)} \chi(\theta) \cos
((n-m)\theta ) d\theta
$$
Clearly,
$$
\|I^1\|_{B(1,\infty)} \leq \frac{1}{2} \sup_{n \in \mathbb{Z}}
\left| \int_{-\pi}^{\pi}  e^{i t (2-2\cos\theta)} \chi(\theta) e^{i
n \theta}d\theta \right| \leq \sup_{a \in \mathbb{R}} \left|
\int_{-\pi}^{\pi}  e^{i t (2-2\cos \theta - a\theta)}
\chi(\theta)  d\theta \right|.
$$
The last expression has been shown in Theorem 3 of \cite{SK} to decay like
$t^{-1/3}$ and this dispersive estimate is sharp. The argument relies
on the van der Corput lemma formulated in Appendix D. Indeed,
if $h(\theta)=2-2\cos(\theta)-a\theta$,
then $h'(\theta_1) = h''(\theta_1)=0$ and $h'''(\theta_1) = 4$ for $a=2$
and $\theta_1=\pi/2$, such that the van der Corput lemma can be applied
with $k = 3$ to produce $t^{-1/3}$ decay.

Proceeding further with $I_V^2$, we have
 \begin{eqnarray*}
 & & (I_V^2 f)_n=\sum_{m,l \in \mathbb{Z}} V_m f_l
\int_{-\pi}^{\pi} e^{i t (2-2\cos \theta )}
\cos(\theta (|n-m|+|m-l|)) \f{\chi(\theta)}
{4 \sin(\theta)} d\theta,
 \end{eqnarray*}
such that
\begin{eqnarray*}
\|I_V^2\|_{B(1,\infty)} & \leq & \sup_{n,l \in \mathbb{Z}} \left| \sum_{m \in \mathbb{Z}}
V_m \int_{-\pi}^{\pi} e^{i t (2-2\cos \theta)-i \theta(|n-m|+|m-l|)}\f{\chi(\theta)}
{4 \sin(\theta)} d\theta \right| \\
& \leq & \|V\|_{l^1}\sup_{a \in \mathbb{R}}
\left| \int_{-\pi}^{\pi} e^{i t (2-2\cos \theta - a \theta)}\f{\chi(\theta)}
{4 \sin(\theta)} d\theta \right|.
\end{eqnarray*}
Thus, we can apply again the van der Corput lemma with
$$
h(\theta)=2-2\cos \theta - a \theta, \quad
g(\theta) = \frac{\chi(\theta)}{4 \sin \theta}.
$$
Since $\chi$ is supported away from $0, -\pi, \pi$, the function $g(\theta)$
is smooth and vanishes in a neighborhood of the end points $-\pi, \pi$.
On the other hand, the function $h(\theta)$ is the same as in the estimate $I^1$.

Finally, we deal with $I_V^3$. We claim first that for all $\si > 5/2$,
\begin{equation}
\label{eq:78}
\sup_{\theta\in[-\pi, \pi]} \sum_m \left|G_m(\theta)\right|+\left|\f{d}{d \theta} G_m(\theta)\right|\leq C
\|V\|_{l^2_{\si}}^2 \|f\|_{l^1}
\end{equation}
We will work with the derivative only, since the estimates for
$G_m(\theta)$ are similar. We have
\begin{eqnarray*}
& &
\f{d}{d \theta}G_m(\theta)=(V \f{d}{d \theta}[ R(2-2\cos \theta)]V f)_m =
2 V_m  \sin \theta R'(2-2\cos \theta )[V f]_m.
\end{eqnarray*}
By \eqref{eq:77} for $j=1$, we obtain for every $\theta\in[-\pi, \pi]$,
\begin{eqnarray*}
\sum_m \left|\f{d}{d \theta} G_m(\theta)\right| & \leq & C|\sin \theta|
\| V \|_{l^2_{\si}} \|R'(2-2\cos \theta)[V f]\|_{l^2_{-\si}} \\
& \leq & C \|V\|_{l^2_{\si}} \f{|\sin \theta|}{\sqrt{2-2\cos \theta}} \|V
f\|_{l^2_\si}\leq C \|V\|_{l^2_{\si}}^2 \|f\|_{l^1}.
\end{eqnarray*}
This finishes the proof of the claim \eqref{eq:78}. Thus, we write
\begin{eqnarray*}
& & (I_V^3 f)_n= \sum_{l \in \mathbb{Z}} f_l \sum_{m \in \mathbb{Z}}
\int_{-\pi}^{\pi} e^{i t (2-2\cos \theta} \cos(\theta (|n-m|+|m-l|))
\f{\chi(\theta) G_m(\theta)}{4 \sin(\theta)}  d\theta,
\end{eqnarray*}
such that
\begin{eqnarray*}
\|I_V^3\|_{B(1,\infty)} & \leq & C \sup_{l \in \mathbb{Z}}
\sum_{m \in \mathbb{Z}}
\left| \int_{-\pi}^\pi e^{i t (2-2\cos \theta)}
e^{-i\theta(|n-m|+|m-l|)}
\f{\chi(\theta) G_m(\theta)}{4 \sin(\theta)}  d\theta \right|
\end{eqnarray*}
We write
$$
h(\theta) = 2 - 2 \cos \theta - \theta a_{t,n,m,l}, \quad
g_m(\theta)=\f{\chi(\theta) }{\sin \theta }G^+_m(\theta),
$$
where $a_{t,n,m,l}=(|n-m|+|m-l|)/t$. Our aim is to estimate
$$
\sum_{m \in \mathbb{Z}} \left| \int_{-\pi}^\pi e^{i t h(\theta)} g_m(\theta) d\theta \right|,
$$
where $g_m(\theta)$ vanishes in a neighborhood of the endpoints
$-\pi, \pi$ and $h(\theta)$ has the property
$$
\max(|h'(\theta)|, |h''(\theta)|, |h'''(\theta)|)\geq 1
$$
as discussed earlier. This is valid for every fixed $t,n,m,l$. We can therefore
apply the van der Corput lemma from Appendix D with
either $k=1,2,3$. In the worst possible scenario, that is $k=3$,
we obtain
$$
\sum_{m \in \mathbb{Z}} \left| \int_{-\pi}^\pi e^{i t h(\theta)} g_m(\theta)
d\theta \right| \leq C t^{-1/3} \sum_{m \in \mathbb{Z}}
\int_{-\pi}^\pi \left| \f{d}{d\theta} g_m(\theta) \right| d\theta \leq C t^{-1/3}
\|V\|_{l^2_{\si}}^2 \|f\|_{l^1},
$$
where the last inequality follows from \eqref{eq:78}.
This finishes the proof of Lemma \ref{le:8}.
\end{proof}

\appendix
\section{Approximation of compact operators}

Here we will prove a lemma, which is used in the proof of Theorem \ref{theo:1}.

\renewcommand{\thelemma}{A}
\begin{lemma}
\label{le:ap1}
Let $X$ be a Banach space and $\{ K_n \}_{n \in \mathbb{N}} : X \to X$
be a sequence of compact operators, such that $\lim_{n \to \infty}
\| K_n-K\|_{B(X,X)} = 0$ for some $K : X \to X$. Then for every $\la\neq 0$,
such that $\la\in \sigma(K)$ with an eigenvector $f \neq 0$, such that
$K f=\la f$, there exists a subsequence $\{ \la_{n_j} \}_{j \in \mathbb{N}}$
of eigenvalues with eigenvectors $\{ f_{n_j}\}_{j \in \mathbb{N}}$, such that
$K_{n_j} f_{n_j} = \la_{n_j} f_{n_j}$ such that $\lim_{j \to \infty} \la_{n_j}
= \la$ and $\lim_{j \to \infty} \| f_{n_j} - f \|_{B(X,X)} = 0$.
\end{lemma}

\begin{proof}
First, we show the existence of a subsequence of eigenvalues
$\la_{n_j}$ of $K_{n_j}$ that converges to eigenvalue $\la$ of $K$.
Then, we construct eigenvectors $f_{n_j}$. Assume the
contrary, that is there exists $\ve_0>0$, so that
$$
0<\ve_0\leq \limsup_{n \to \infty} {\rm dist}(\si(K_n), \la).
$$
By the functional calculus,
there exists a subsequence $\{ K_{n_j} \}_{j \in \mathbb{N}}$, such that
$$
\|(K_{n_j}-\la)^{-1}\|_{B(X,X)}\leq 2\ve_0^{-1}
$$
Pick any eigenvector $f$ for the eigenvalue $\la$, such that $Kf=\la f$
and $\|f\|_X=1$.
Let $g_j=(K_{n_j}-\la)f=(K_{n_j}-K)f$. Clearly, $\|g_j\|_X\leq
\|K_{n_j}-K\|_{B(X,X)}\to 0$. On the other hand,
$$
1=\|f\|_X=\|(K_{n_j}-\la)^{-1} g_j\|_X\leq 2\ve_0^{-1} \|g_j\|_X\to 0.
$$
A contradiction arises, whence there is a subsequence
$\{\la_{n_j}\}_{j \in \mathbb{N}}$ which converges to $\la$.  Pick
eigenvectors $f_{n_j}$, such that $K_{n_j} f_{n_j}=\la_{n_j} f_{n_j}$ and
$\|f_{n_j}\|_X=1$. Since $K$ is compact, it follows
that $K f_{n_j}$ will  have a convergent subsequence, call it again $f_{n_j}$.
Let $g:=\lim_{j\to \infty} K f_{n_j}$. We have
\begin{eqnarray*}
\|g-\la f_{n_j}\|_X &\leq &
|\la-\la_{n_j}|+ \|Kf_{n_j}-g\|_{X}+
\|(K-K_{n_j})f_{n_j}\|_X \leq \\
 &\leq& |\la-\la_{n_j}|+
\|Kf_{n_j}-g\|_{X}+\|(K-K_{n_j})\|_{B(X,X)}\to 0.
\end{eqnarray*}
Thus, the subsequence $\{ f_{n_j} \}_{j \in \mathbb{N}}$ converges to
$f:=g/\la$ in $B(X,X)$ norm if $\la \neq 0$.
Also, $\la f = g = \lim_{j \to \infty} K f_{n_j} = Kf$.
\end{proof}

\section{Conditions on generic potentials}

Let us consider the difference equation $(-\Delta + V) \psi = 0$ or
\begin{equation}
\label{solution-psi} \psi_{n+1} + \psi_{n-1} = (2 + V_n) \psi_n,
\qquad n \in \mathbb{Z}.
\end{equation}
Two fundamental solutions of (\ref{solution-psi}) are defined by the
discrete Green function in the form
\begin{eqnarray*}
\psi^+_n & = & 1 - \sum_{m = n}^{\infty} (n-m) V_m
\psi^+_m, \\
\psi^-_n & = & 1 + \sum_{m = -\infty}^n (n-m) V_m \psi^-_m.
\end{eqnarray*}
It is straightforward to check that the discrete Wronskian
$$
W[\psi^+,\psi^-] = \psi_{n+1}^- \psi_n^+ - \psi_{n+1}^+ \psi_n^-
$$
is constant on $n \in \mathbb{Z}$. Therefore, the Green function
representation of $\psi^+$ and $\psi^-$ immediately implies
that $\sum_{m \in \mathbb{Z}} V_m \psi_m^+ = \sum_{m \in \mathbb{Z}} V_m
\psi_m^-$ or simply $\langle V, \psi^+ \rangle = \langle V,
\psi^- \rangle$. If $V \in l^1_1$, then
$$
\lim_{n \to +\infty} \psi_n^+ = \lim_{n \to -\infty}
\psi_n^- = 1
$$
and
$$
- \lim_{n \to -\infty} \frac{\psi_n^+}{n} = \lim_{n \to +\infty}
\frac{\psi_n^-}{n} = \langle V, \psi^+ \rangle = \langle V,
\psi^- \rangle.
$$
It follows by this construction that the solution of $(-\Delta + V) \psi_0 = 0$
spanned by the fundamental solutions $\psi^+$ and $\psi^-$ always
exists in $l^2_{-\sigma}$ for $\sigma > \frac{3}{2}$ (Remark \ref{remark-potential}).
We can now prove the equivalence of conditions in Definition \ref{definition-generic-potential}
and the conditions in Theorem \ref{theo:2}.

\renewcommand{\thelemma}{B}
\begin{lemma}
Let $V \in l^1_1$. The two conditions are equivalent:
\begin{enumerate}
\item No solution
$\psi_0$ of equation $(-\Delta + V) \psi_0 = 0$ exists in
$l^2_{-\sigma}$ for $\frac{1}{2} < \sigma \leq \frac{3}{2}$

\item No solution $u_0$ of equation $T_0 u_0
= 0$ exists in $l^2_{-\si}$ for any $\sigma > \frac{3}{2}$ and
$\langle V, T_0^{-1} 1 \rangle \neq 0$
\end{enumerate}
\label{lemma-equivalence}
\end{lemma}

\begin{proof}
Condition 1 is equivalent to the constraint $\langle V, \psi^+ \rangle
\neq 0$. Indeed, if $\langle V, \psi^+ \rangle = 0$, then $\psi^+ \in
l^{\infty}(\mathbb{Z})$ and thus $\psi^+ \in l^2_{-\si}$ for
$\si > \frac{1}{2}$. If $\langle V, \psi^+ \rangle \neq 0$, then
no solution $\psi$ of equation (\ref{solution-psi}) exists in
$l^2_{-\si}$ for $\frac{1}{2} < \si \leq \frac{3}{2}$.

Let $u$ be a solution of $T_0 u = 1$, which can be rewritten in the
explicit form
\begin{equation}
\label{solution-u} u_n = 1 + \frac{1}{2} \sum_{m = -\infty}^n (n-m)
V_m u_m - \frac{1}{2} \sum_{m = n}^{\infty} (n-m) V_m u_m.
\end{equation}
Direct computations show that $u$ solves the same difference
equation (\ref{solution-psi}) and
$$
\lim_{n \to +\infty} \frac{u_n}{n} = -\lim_{n \to -\infty}
\frac{u_n}{n} = \langle V, u \rangle.
$$
Therefore, $u = c( \psi^+ + \psi^-)$ with $c \neq 0$ and the
constraint $\langle V, u \rangle = \langle V, T_0^{-1} 1 \rangle
\neq 0$ is equivalent to the constraint $\langle V, \psi^+
\rangle \neq 0$ that is condition 1.

Assume now that there exists a solution of equation $T_0 u_0 = 0$ in
$l^2_{-\si}$ for $\si > \frac{3}{2}$. This function is a solution of
the same equation (\ref{solution-u}) but without the constant term
on the right-hand-side. Therefore, $u_0$ satisfies
(\ref{solution-psi}) and $u_0$ is linearly independent from $u$,
which is another solution of (\ref{solution-psi}). Multiplying
equation (\ref{solution-u}) by $V_n (u_0)_n$ and summing over $n \in
\mathbb{Z}$, we obtain
$$
\langle V, u_0 \rangle = \sum_{m \in \mathbb{Z}} V_m u_m \left[
(u_0)_m - \frac{1}{2} \sum_{n \in \mathbb{Z}} |m-n| V_m (u_0)_m
\right] = 0.
$$
Therefore, $u_0 \in l^2_{\sigma}$ for $\sigma > \frac{1}{2}$ for
exponentially decaying potentials $V$, that is there exists $\psi_0$
of equation (\ref{solution-psi}) in $l^2_{\sigma}$ for $\sigma >
\frac{1}{2}$. In the opposite direction, if there exists $\psi_0$ of
equation (\ref{solution-psi}) in $l^2_{\sigma}$ for $\sigma >
\frac{1}{2}$, then $\psi_0 = c \psi^+$ with $c \neq 0$ and,
since $\langle V, \psi^+ \rangle = 0$ in this case, one can
choose $c = \frac{1}{2} \sum_{m \in \mathbb{Z}} m V_m (\psi_0)_m$ so
that $\psi_0$ solves $T_0 \psi_0 = 0$, that is there exists $u_0 =
\psi_0$. Therefore, existence of $\psi_0$ is equivalent to existence
of $u_0$. Combining both results together, we have established the
equivalence of conditions 1 and 2.
\end{proof}

\section{Jensen--Kato Lemma}

A general lemma to estimate the oscillatory integrals  is provided in \cite{JK}.
Here, we formulate and prove a simplified version, which is used in the proof
of Theorem \ref{theo:dec1}.

\begin{lemma}
\label{le:90}
Let  $\cb$ be a Banach space, so that for $F\in C^2(0,a; \cb)$,
 $F(0)=F(a)=0$, \\
$\|F'(\om)\|_{\cb}\leq C
\om^{-1/2}$ and $\|F''(\om)\|_{\cb}\leq C \om^{-3/2}$ as $\om\to 0$. Then for every $t>1$,
\begin{equation}
\label{eq:89}
\left\| \int_0^a e^{i t w} F(\om) d\om  \right\|_{\cb}\leq C  |t|^{-3/2}.
\end{equation}
\end{lemma}
\begin{proof}
Take an unit element $b^*$ in the dual space $B^*$. Then, it is clearly enough to show
$$
\left| \int_0^a e^{i t w} b^*[F(\om)] d\om  \right| \leq C |t|^{-3/2}.
$$
where $C$ is independent of $B^*$. Thus, \eqref{eq:89} follows from its scalar version, where
$\tilde{F}(\omega):=b^*[F(w)]$, since the estimates for $F$ carry over $\tilde{F}$. That is without
loss of generality, we may assume that $\cb={\mathbf R}$.

Next, an integration by parts yields
$$
 \int_0^a e^{i t w} F(\om) d\om=\f{i}{t} \int_0^a e^{i t w} F'(\om) d\om= \f{i}{t} \left[
 \int_0^{\min(1/|t|, a)} e^{i t w} F'(\om) d\om+ \int_{\min(1/|t|, a)}^a e^{i t w} F'(\om) d\om\right]
$$
For the first integral, estimate by absolute value and $|F'(\om)|\leq C |\om|^{-1/2}$,
$$
C \f{1}{|t|} \int_0^{\min(1/|t|, a)} |\om|^{-1/2} d\om\leq 2C |t|^{-3/2}.
$$
For the second integral, perform one more integration by parts and $|F''(\om)|\leq C \om^{-3/2}$,
\begin{eqnarray*}
& & \f{1}{|t|}\left|\int_{\min(1/|t|, a)}^a e^{i t w} F'(\om) d\om\right|\leq
\f{C}{|t|^2} \left[|F'(\min(1/|t|, a))|+  \int_{\min(1/|t|, a)}^a |\om|^{-3/2} d\om\right]\leq C_1
|t|^{-3/2}.
\end{eqnarray*}
These bounds complete the proof of the lemma.

\end{proof}

\section{Estimation of oscillatory integrals}

The van der Corput lemma is stated as a corollary on page 334 in \cite{Stein}. \\

{\bf Van der Corput Lemma:} {\em Suppose $\phi$ is a real-valued function, smooth in $(a,b)$,
so that $|\phi^{(k)}(x)|\geq 1$ for some integer $k$.
(If $k=1$, we will have to also assume that $\phi'(x)$ is monotonic). Then,}
\begin{equation}
\label{eq:180}
\left| \int_a^b e^{i \la \phi(x)} \psi(x) dx \right| \leq
c_k \la^{-1/k} \left[\psi(b)+\int_a^b |\psi'(x)|
dx\right].
\end{equation}

\medskip

Here we will prove a lemma which is used in the proof of Lemma \ref{le:7}. This
lemma is basically a corollary of the van der Corput lemma.

\renewcommand{\thelemma}{D}
\begin{lemma}
\label{le:91}
Assume that the function $g(\theta)$ is continuously differentiable in
$[-\theta_0, \theta_0]$ for certain $0 < \theta_0 < \frac{\pi}{4}$
and  $\sup_{-\theta_0\leq \theta\leq \theta_0} [|g'(\theta)|+ |g(\theta)|]\leq C$. Then
\begin{equation}
\label{eq:780}
\sup_{a \in \mathbb{R}} \left|\int e^{it(2-2\cos \theta - a\theta)} \chi_0(\theta)
g(\theta)
d\theta\right|\leq
C t^{-1/2}
\end{equation}
for any $t > 0$.
\end{lemma}

\begin{proof}
To use the van der Corput lemma, we need to check that for a fixed parameter
$a$ and on the support of the function $\chi_0$, the phase function $h(\theta) = 2 - 2 \cos \theta - a \theta$ satisfies the condition that $\max(|h'(\theta)|, |h''(\theta)|)\geq 1$ for every fixed $\theta$.
Assuming that claim (and observing that $\theta \to h'(\theta) = 4\sin \theta - a$
is a monotonic function),  we may apply the
van der Corput lemma with either $k=1$ or $k=2$, which gives us \eqref{eq:780}. Thus,
compute $h'(\theta) = 4 \sin \theta - a$ and $h''(\theta) = 4 \cos \theta$,
whence
$$
(h')^2 + (h'')^2 = 16 - 8 a \sin \theta + a^2 \geq 8,
$$
where the last inequality is a consequence of $|\sin \theta | \leq 1/\sqrt{2}$ in the
interval under consideration. Therefore, $\max(|h'(\theta)|, |h''(\theta)|)\geq
2\sqrt{2}>1$.
\end{proof}

\end{document}